\documentclass[aps,prl,twocolumn,showpacs,preprintnumbers,superscriptaddress]{revtex4}
\usepackage{graphicx}
\usepackage{dcolumn}
\usepackage{bm}

\begin{document}

\title{Anderson localization in carbon nanotubes: defect density and temperature effects}

\author{Blanca Biel}
\affiliation{
Departamento de F\'\i sica Te\'orica de la Materia Condensada, Universidad Aut\'onoma de Madrid, E-28049 Madrid, Spain
}
\author{F.J. Garc\'\i a-Vidal}
\affiliation{
Departamento de F\'\i sica Te\'orica de la Materia Condensada,
Universidad Aut\'onoma de Madrid, E-28049 Madrid, Spain }

\author{Angel Rubio}
\affiliation{Institut f\"ur Theoretische Physik, Freie
Universit\"at Berlin, Arnimallee 14, D-14195 Berlin, Germany}
\affiliation{Departamento de F\'{\i}sica de Materiales, Facultad
de Ciencias Qu\'{\i}micas, UPV/EHU, Centro Mixto CSIC-UPV/EHU}
\affiliation{Donostia International Physics Center, E-20018 San
Sebasti\'an, Spain} \affiliation{European Theoretical Spectroscopy
Facility}

\author{Fernando Flores}
\affiliation{
Departamento de F\'\i sica Te\'orica de la Materia Condensada, Universidad Aut\'onoma de Madrid, E-28049 Madrid, Spain
}
\date{\today}

\begin{abstract}
The role of irradiation induced defects and temperature in the
conducting properties of single-walled (10,10) carbon nanotubes
has been analyzed by means of a first-principles approach. We find
that di-vacancies modify strongly the energy dependence of the
differential conductance, reducing also the number of contributing
channels from two (ideal) to one. A small number of di-vacancies
(5-9) brings up strong Anderson localization effects and a seemly
universal curve for the resistance as a function of the number of
defects. It is also shown that low temperatures, around $15-65$K,
are enough to smooth out the fluctuations of the conductance
without destroying the exponential dependence of the resistivity
as a function of the tube length.
\end{abstract}


\maketitle


The ubiquity of defects in materials science limits somehow the
performance of the material; this is even more dramatic when the
dimensionality of the system is reduced moving towards
nanostructures, in particular carbon nanotubes \cite{book}.
Understanding and controlling the conductance of these systems is
decisive for their possible application in molecular devices
\cite{terrones,marco,trans0,trans,banhart,jkong,nmat,kras}. In
perfect single-walled carbon nanotubes (SWNTs), electrons
propagate ballistically if the inelastic processes can be
neglected, i.e. if the electronic phase coherence length,
$L_{\phi}$, is larger than the nanotube length, $L$. High quality
metallic SWNTs exhibit $L_\phi$ as large as one micron
\cite{trans0}. If inelastic processes, like the electron-phonon
interaction, are important, the system conducts within the
diffusive regime. On the other hand, it has been shown
\cite{jkong,nmat} that defects induce Anderson localization in the
electron states of a SWNT if $L_0<L<L_{\phi}$, $L_0$ being the
localization length. This is the regime we are interested in.

This Anderson localization phenomenon is due to the concerted
action of quantum confinement and defect scattering effects
\cite{pendry,datta}. The theoretical calculations presented in
Ref. \cite{nmat} were concentrated on the zero-temperature case.
Temperature (T) adds new de-phasing mechanisms for the coherent
electron transport in SWNTs that should be considered on the same
footing as the other sources of de-phasing. In this context, a
fully microscopic study of the electronic transport in defected
SWNT in the Anderson localization regime taking into account the
effect of finite T is lacking. This is the goal of the present
Letter: to fully understand the properties of the localization
regime as a function of the nanowire length, temperature, density
of defects and strength of the defect scattering potential for the
paradigmatic (10,10)-chirality.


As we are interested in modelling the role of defects in the
conductivity of SWNTs, we have chosen to analyze in detail the
mono-vacancies and di-vacancies that are the most common defects
that appear upon atom irradiation \cite{kras}; the comparison of
these two cases will allow us to analyze the electron localization
and its effect on the nanotube conductance as a function of the
different scattering potentials associated with those defects. As
our calculations of the zero-bias conductance drop associated with
a single defect show a conductance of $1.98G_0$ for the
mono-vacancy and $1.36G_0$ for the lateral di-vacancy (
$2G_0=4e^2/h$ is the conductance of an ideal nanotube),
di-vacancies will show a more dramatic effect in the nanotube
conductance \cite{nmat}. Accordingly, although we will present
results for mono-vacancies and di-vacancies in order to draw more
general conclusions, most of our discussion will be concentrated
on the di-vacancy case.


The conductance calculations are done within a first-principles
Local Orbital Density Functional (LODF) method that maps the full
hamiltonian into a local-orbital one, allowing the use of the
standard machinery of non-equilibrium Green´s function in order to
extract the transmission probability, and from that the nanotube
conductance \cite{FJ04}. The simulation geometry consists of a
device region with the defected nanotubes connected to two
semi-infinite perfect tubes. The geometry of the nanotube around
each vacancy is calculated using this LODF method and the defected
nanotube is defined by adding $N$ defects separated by a mean
distance $d$. Then, the length of the nanotube, $L$, is $L=N
\times d$. The defects are assumed to be distributed along the rim
of the tube with a distance between consecutive defects (measured
along the tube axis) presenting a uniform random distribution
between $0$ and $2d$.  The conductance calculations are performed
for a statistically significant number of defect realizations. For
the sake of simplicity, we present results for a nanotube with
either only lateral di-vacancies (this is the most favorable
geometry for di-vacancies \cite{nmat}) or with only
mono-vacancies. In our calculations, we obtain an effective local
orbital hamiltonian associated with the sp$^3$-basis set of the
FIREBALL-orbitals \cite{fireball}; details about the technique can
be found in Ref.\cite{nmat}. Due to the local nature of this
approach, it is feasible to calculate the electronic properties of
very long nanotubes (up to several microns long) with an arbitrary
distribution of defects.

First we analyze the energy dependence of the nanotube
differential conductance, $g(E)$, for two different density of
defects and zero T. In Fig.1 we plot $g(E)$ of a (10,10) carbon
nanotube for four random configurations of defects with two
different inter-defect distances ($d=45.4$nm and $75.5$nm) and two
number of defects ($15$ and $25$) introduced in the tube, typical
values observed in the experiments \cite{nmat}. In all cases we
observe strong fluctuations of the differential conductance as a
function of energy: this is a clear indication of the localized
nature of the electronic states. Notice also, comparing panels (a)
and (b) of Fig. 1, that $g(E)$ also depends strongly on $d$. For
finite T calculations the crucial point to realize is that
electrons are injected into the nanotube with energies within a
window of the order of $k_BT$. Then, it is expected that the
strong fluctuations in the conductance would be washed out for a
finite T. This will happen if the thermal coherent length $L_T$
($L_T=\hbar v_F/k_BT$, $v_F$ being the Fermi velocity) is smaller
than $L_0$. One naively say also that Anderson localization could
not appear in this limit. This would be the case only if
$L_{\phi}$ is of the order of $L_0$. As perfect metallic tubes
exhibit ballistic response over long distances (microns), the
electrons injected in the nanotube with different energies do not
suffer inelastic scattering events and they behave as {\it
independent electrons} \cite{fonones}. In this picture, the total
conductance is just the sum of many contributions from electrons
injected within the thermal energy window \cite{quantumdots}.

\begin{figure}[h]\label{g(e)}
\includegraphics[width=0.9\columnwidth]{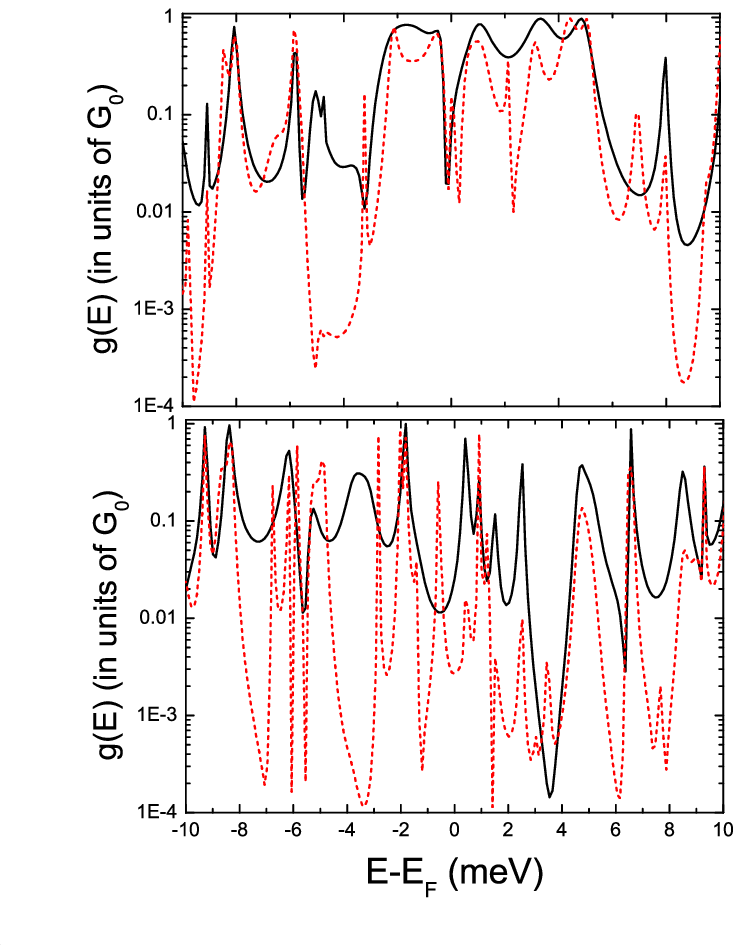}
\caption{ (color on line) Differential conductance for two
defected single-wall carbon nanotubes. Top panel, corresponds to a
distance between defects of $d=45.4$nm and the bottom panel for
$d=75.5$nm. Two different number of defects (15, continuous line,
and 25, dashed line) are analyzed in both panels. Notice that the
plotted energy window is smaller than the room temperature energy
window ($25$meV).}
\end{figure}

In Fig. 2a we show the dependence of the nanotube resistance with
the tube length for different temperatures and a given random
distribution of di-vacancies. These curves have been calculated as
follows. First, we evaluate the resistance of the longest tube
(corresponding to $N=25$ defects in this case) with a given random
configuration of di-vacancies. Next, we calculate the resistance
of a nanotube with $N-1$ defects just eliminating the last
di-vacancy (the length of the nanotube is then reduced to
$L=(N-1)d$). This procedure is repeated until we end with a
nanotube containing only one di-vacancy. Then, the whole
calculation is repeated for several temperatures but with the same
random configuration of defects. At zero T, as expected, the
fluctuations in the resistance are very strong. The effect of
temperature is striking: at room T, there is a complete reduction
of the random fluctuations seen at $T = 0$K. However we still get
the exponential increase of resistance with tube length that
points out that the Anderson localization regime survives even at
room T. Figure 2a also demonstrates that only a very low T (less
than $20$K) is needed to wash out the fluctuations of the
resistance as a function of the tube length. This result explains
why fluctuations are not seen at room T as experimentally reported
in \cite{nmat}. In Fig. 2b we plot both the zero and room T
resistances for several random configurations of di-vacancies with
$d = 75.5$nm. Notice that at room T the resistance of different
configurations is very insensitive to the particular random
distribution of di-vacancies. This is not the case for $T=0$K
where there are strong fluctuations with respect to the average
value.

\begin{figure}[h]\label{fig2.eps}
\includegraphics[width=0.9\columnwidth]{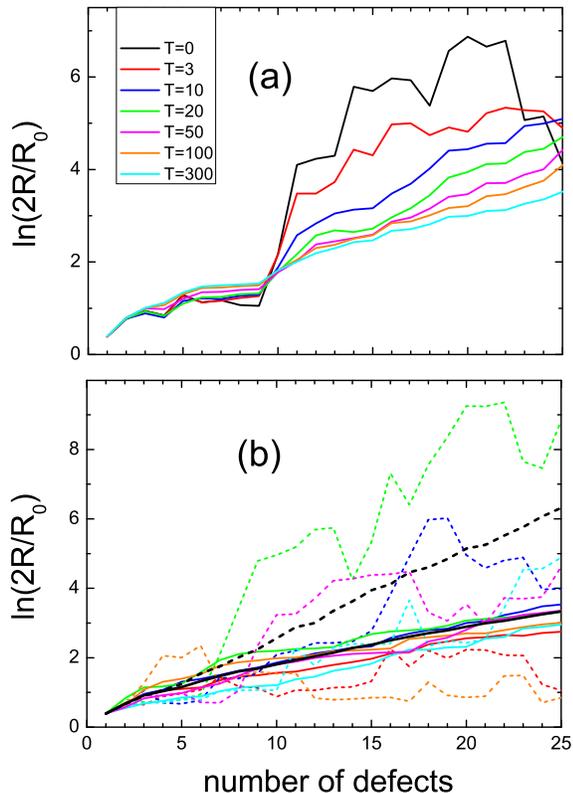}
\caption{ (color on line) a) Temperature dependence of the
calculated resistance for a defected nanotube with $d=75.5$nm. b)
For T=0 (dashed lines) and T$=300$K (continuous curves) we plot
the calculated resistances for different random defect
configurations with $d=75.5$nm. The mean value of the resistances
is also plotted in black.}
\end{figure}

Figure 3 shows our calculations for the mean value of the room T
resistance (as a result of an average over 15 random cases, as
explained above) as a function of the carbon nanotube length for
different $d$'s. Our results indicate that this room T resistance
seems to fit a universal curve once it is plotted in terms of the
number of defects ($N$) instead of the total length ($L$). An
important result of performing a finite T calculation is that the
resistance curves of Fig.3 fit better as $R_0  \times \exp(L/L_0)$
than as $R_0 /2 \times \exp(L/L_0)$ ($R_0$ being $1/G_0$), for
$N>3-5$. This is a strong indication of the reduction of the
conducting channels induced by both defects and finite T
\cite{frank,sanvito}; only one channel is contributing to the
total conductance. Moreover, this also suggests that the nanotube
enters in the localization regime for a very low density of
defects, i.e. with more than 5-9 di-vacancies, provided the
resulting $L_0$ is smaller than $L_{\phi}$. We have looked into
the origin of the reduction in the number of channels in our
system by diagonalizing the transmission matrix \cite{pavel} used
in our calculations. The diagonalization supports the previous
discussion, although the full reduction to one-channel only
appears once 5-9 defects have been introduced in the nanotube.
Notice that Figs. 1 and 2 already accounted for this result: in
Fig.1, for both $25$ or $15$ defects the differential conductance
is never larger than $G_0$; while in Fig. 2, the resistance in
this case is never smaller than $R_0$ for a number of defects
larger than 5.

The inset of Fig. 3 shows the effective localization length,
$L_0$, as a function of $d$ for di-vacancies at room T indicating
that $L_0 \approx 8.5d$. In Ref.\cite{nmat} we found $L_0 \approx
4.1d$ after averaging the resistance in the log-scale over many
different random cases; note that this is appropriate because of
the normal distribution of $log(R)$ in 1D systems \cite{pendry}.
Our new localization length reflects, however, that at room T, and
for a particular distribution of di-vacancies, the conductance of
the system is calculated as the sum of the channel contributions
associated with the energy window $E_F \pm k_BT$ (which represents
a kind of effective average on energy of the conductance). This
new estimation of $L_0$ at room T slightly modify the average
distance between di-vacancies that were estimated in Ref.
\cite{nmat} to have been created in the SWNTs by Ar$^+$,
suggesting that those values have to be reduced in all the cases
by around $50 \%$. Notice that this density of di-vacancies was
obtained neglecting the effects that mono-vacancies are producing
in the nanotube conductance. This is supported by our calculations
for mono-vacancies whose results can be summarized in the
following way : (i) the ballistic regime is found to extend up to
around 200 mono-vacancies ; (ii) the localization regime appearing
for more than $200$ defects shows a resistance as a function of
the tube length that can also be fitted to $R_0 \times exp(L/L_0)$
with $L_0 \approx 600d$, implying that in the localization regime
the conductance is again controlled by only one channel.
Therefore, mono-vacancies show a much smaller effect on the
nanotube conductance than di-vacancies do. It is expected that for
large vacancy clusters (larger than di-vacancies), $L_0$ would be
also proportional to $d$, but with a proportionality constant much
smaller than $8.5$.

\begin{figure}[h]\label{fig3_new.eps}
\includegraphics[width=\columnwidth]{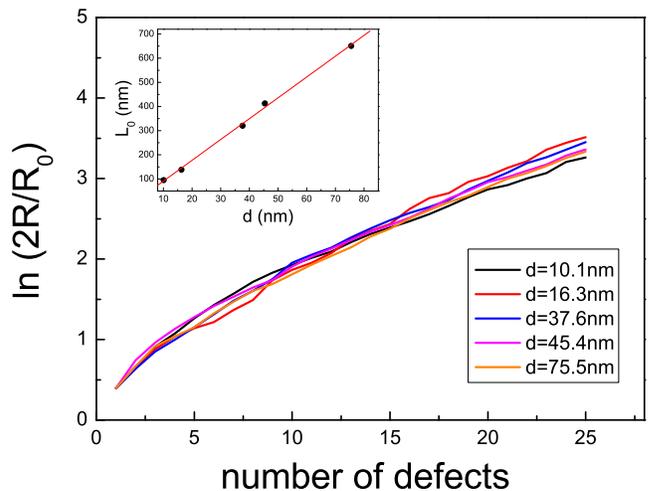}
\caption{ (color on line) Calculated room temperature resistance
for different average distance between di-vacancies ($d$) as a
function of number of defects. Inset: localization length, $L_0$,
extracted from the fitting $R=R_0  \times \exp(L/L_0)$ as a
function of $d$.}
\end{figure}

Finally for the di-vacancy case we have analyzed the effect of
temperature in the quenching of the conductance fluctuations. This
has been done for a tube with 25 di-vacancies as follows. As we
are interested in the fluctuations that have already disappeared
at room T, we take the calculated $R(L)$ at room T as reference
data. The scaled resistance plot is shown in the inset of Fig.4
for $d=75.5$nm for a particular random configuration (the same as
analyzed in Fig.2a), after substraction of its mean value in $L$.
We perform this calculation for all random configurations of
defects and calculate for each of them the r.m.s deviation
($\sigma$) with respect to the $R(L)$ data at room T. The final
average over all configurations of $\sigma$ is plotted in Fig.4.
The general trend is as follows: the higher T the faster the
fluctuation damping. Also, the fluctuations decrease more rapidly
for a low density of defects (large $d$). Figure 2a shows that for
$d=75.5$nm, the fluctuations are small for T higher than T$_C
\approx 15$K, whereas for the other two $d$'s (not shown here)
this occurs for T$_C \approx 32$K and T$_C \approx 65$K, for
$d=37.6$nm and $16.3$nm, respectively. Thus, the temperature at
which the fluctuations are quenched scales as $1/d$. This can be
understood by noting that at T$_C$ the thermal length $L_T$ should
be of the order of the localization length, $L_0$ (that is
proportional to $d$). Physically, the smaller the density of
defects the smaller the averaged scattering potential, therefore a
lower T would be needed to wash out the fluctuations. We have also
carried out similar numerical calculations for (5,5) metallic
SWNTs (not shown here). The underlying physics in this case seems
to be the same than the one discussed for (10,10) SWNTs, allowing
us to safely conclude that our theoretical findings would remain
valid for metallic SWNTs of different chirality.

\begin{figure}[h]\label{fig4.eps}
\vspace*{0.3 cm}
\includegraphics[width=\columnwidth]{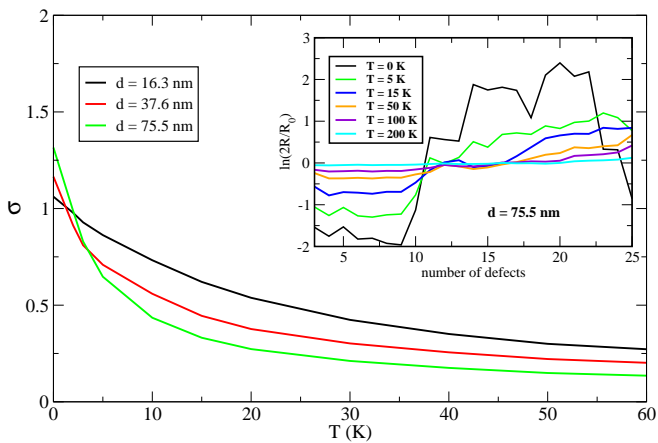}
\caption{ (color on line) Calculated r.m.s deviation ($\sigma$) as
a function of temperature for three different density of
di-vacancies. The inset shows the calculated resistance referred
to the room temperature results for a particular random
configuration with $d=75.5$nm (see text for discussion).}
\end{figure}

In conclusion, we have shown that in a (10,10) carbon nanotube:
(i) the transition between the ballistic and the localization
regimes appears for a small number of di-vacancies in the nanotube
(around $3-5$). (ii) For a higher number of defects the system
shows localization, reducing the number of effective channels from
two (ballistic) to one. (iii) At zero T, the conductance of the
nanotube as a function of its length shows strong fluctuations.
The net effect of a finite T is to wash out the strong
fluctuations presented at $T=0$K; our calculations show that those
fluctuations could be observed for di-vacancies (with $d$ smaller
than around $70$nm and $N > 5-9$) if the temperature is below
$15-65$K. We stress here that, in spite of the disappearance of
the fluctuations, the exponential behavior of $R(L)$ is still
preserved at room T. This puts in evidence our assumption of a
very low inelastic scattering in SWNTs, i.e. the phase coherence
length of electrons much larger than the localization length.

Our results are important in order to understand, in a further
study, the high bias conductance where optical phonons do play a
key role in limiting the conductance. Still there is a lack of
knowledge about the role played by both temperature and defect
density in this regime.

The authors were partially supported by the Spanish MCyT under
contract MAT2002-01534 and the EC 6th Framework Network of
Excellence NANOQUANTA (NMP4-CT-2004-500198). B.B. is indebted to
MEC (Spain) for a F.P.U. fellowship. Computing time for some of
these calculations in the Centro de Computaci\'on Cient\'ifica de
la UAM is gratefully acknowledged. A.R. acknowledges the Humboldt
Foundation under the Bessel research award (2005). We thank J.
G\'omez-Herrero for a very intensive and fruitful collaboration.


\begin{thebibliography}{99}
\bibitem{fonones} At low bias, acoustic phonons are the only source of scattering
for an otherwise perfect nanotube. The mean-free path is very long
what makes this de-phasing mechanism not relevant for our study in
which the inter-defect distance is much smaller.
\bibitem{quantumdots} The present analysis does not support the model
of a tube consisting of series of quantum-dots. If this were the
case, $L_0$ would be just smaller than $d$ and, as shown in this
paper, this is not the case.
\end{thebibliography}
\end{document}